%% file: template.tex
\documentclass[preprint,journal]{vgtc}       

\ieeedoi{10.1109/TVCG.2020.3030382}

\input{"packages.tex"}
\input{"commands.tex"}
\input{"variables.tex"}

\usepackage{mathptmx}
\usepackage{graphicx}
\usepackage{times}

\usepackage[bookmarks,backref=true,linkcolor=black]{hyperref} 
\hypersetup{
  pdfauthor = {},
  pdftitle = {},
  pdfsubject = {},
  pdfkeywords = {},
  colorlinks=true,
  linkcolor= black,
  citecolor= black,
  pageanchor=true,
  urlcolor = black,
  plainpages = false,
  linktocpage
}

\onlineid{0}

\vgtccategory{Research}

\vgtcinsertpkg

\title{SafetyLens: Visual Data Analysis of Functional Safety of Vehicles}

\author{Arpit Narechania\thanks{e-mail: arpitnarechania@gatech.edu}\\ %
    \scriptsize Georgia Institute of Technology %
    \and Ahsan Qamar \thanks{e-mail: aqamar2@ford.com}\\ %
    \parbox{1.4in}{\scriptsize \centering Ford Motor Company}
    \and Alex Endert\thanks{e-mail: endert@cc.gatech.edu}\\ %
    \scriptsize Georgia Institute of Technology %
}

\shortauthortitle{Narechania et al.}

\abstract{
Modern automobiles have evolved from just being mechanical machines to having full-fledged electronics systems that enhance vehicle dynamics and driver experience. However, these complex hardware and software systems, if not properly designed, can experience failures that can compromise the safety of the vehicle, its occupants, and the surrounding environment. For example, a system to activate the brakes to avoid a collision saves lives when it functions properly, but could lead to tragic outcomes if the brakes were applied in a way that's inconsistent with the design. 
Broadly speaking, the analysis performed to minimize such risks falls into a systems engineering domain called \fusa.
In this paper, we present \app, a visual data analysis tool to assist engineers and analysts in analyzing automotive \fusa datasets. \app combines techniques including network exploration and visual comparison to help analysts perform domain-specific tasks.
This paper presents the design study with \users that resulted in the design guidelines, the tool, and user feedback. 

} 


\CCScatlist{ 
 \CCScat{K.6.1}{Management of Computing and Information Systems}%
{Project and People Management}{Life Cycle};
 \CCScat{K.7.m}{The Computing Profession}{Miscellaneous}{Ethics}
}

\teaser{
  \includegraphics[width=1\textwidth]{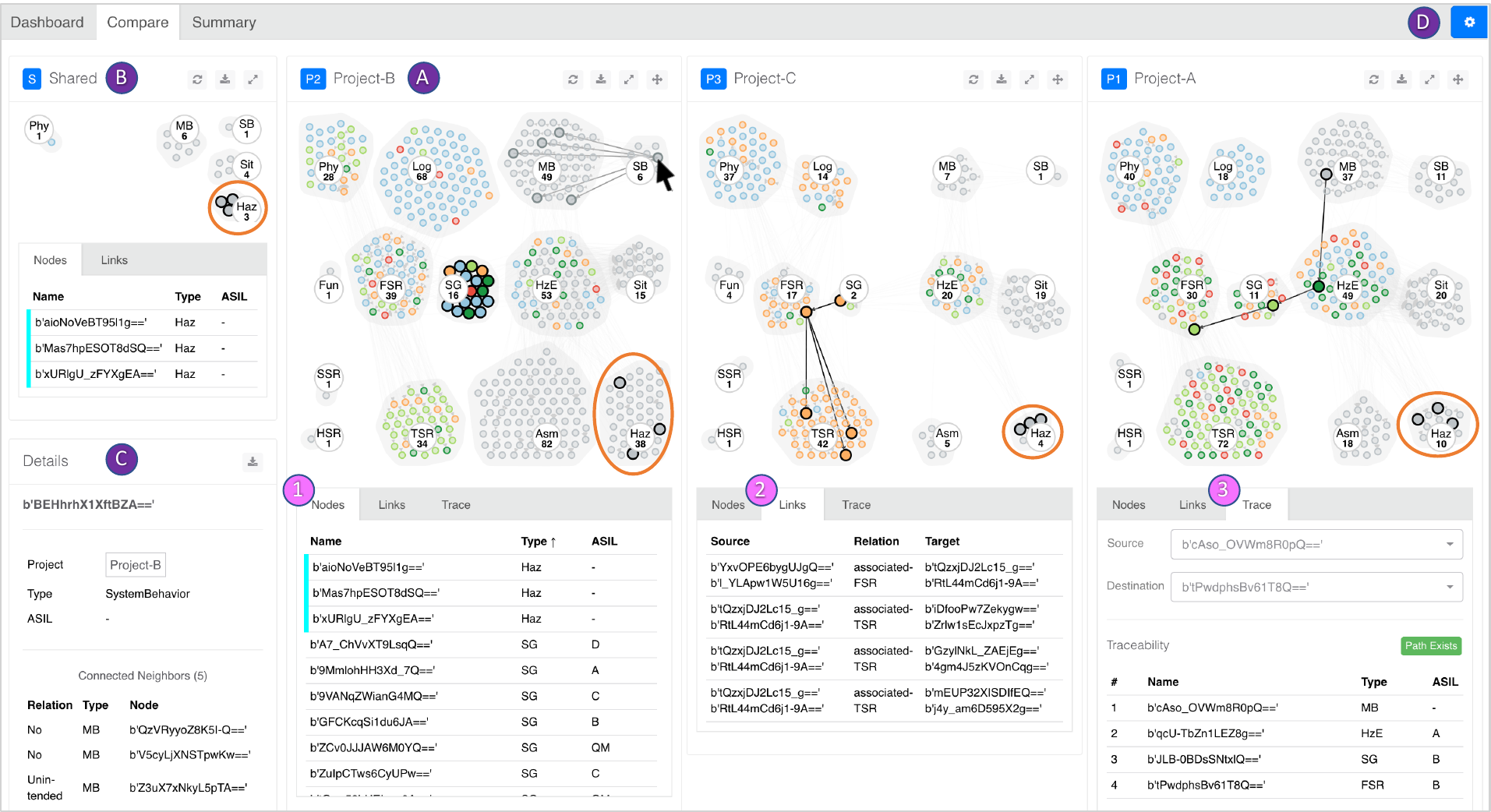}
  \caption{The \app Compare view: \textcolor{purple}{\textbf{(A)}} Project Panel comprising the visualization canvas and three tabs: \textcolor{pink}{\textbf{(1)}} Nodes, \textcolor{pink}{\textbf{(2)}} Links, and \textcolor{pink}{\textbf{(3)}} Trace, \textcolor{purple}{\textbf{(B)}} Shared Nodes Panel with three nodes (\emph{Hazards}) selected and highlighted across all three projects \textcolor{orange}{(ellipses)}, \textcolor{purple}{\textbf{(C)}} Detail View Panel showing details of a node (\textit{System Behavior}) hovered in \emph{Project-B} ({\small{\faMousePointer}}), and \textcolor{purple}{\textbf{(D)}} opens into a Control Panel comprising search and filter controls.
  }
  \label{fig:interface}
 }

\begin{document}


\firstsection{Introduction}

\maketitle
 


\input{"src/sections/introduction.tex"}

\section{Industrial Background}
\label{section:industrial-background}
\input{"src/sections/industrial_background.tex"}

\section{Related Work}
\label{section:related-work}
\input{"src/sections/related_work.tex"}

\section{Design Process and Domain Requirements}
\label{section:domain-exploration}
\input{"src/sections/domain_exploration/main.tex"}

\section{SafetyLens}
\label{section:safetylens}
\input{"src/sections/system_overview.tex"}

\section{Usage Scenarios}
\label{section:usage-scenario}
\input{"src/sections/use_cases.tex"}

\section{Discussion}
\label{section:discussion}

\input{"src/sections/discussion.tex"}



\section{Conclusion}
\label{section:conclusion}
\input{"src/sections/conclusion.tex"}

\input{"src/sections/acknowledgements.tex"}

\bibliographystyle{abbrv}
\bibliography{template}
\end{document}

%% file: packages.tex
\usepackage{microtype}                 
\PassOptionsToPackage{warn}{textcomp}  
\usepackage{textcomp}                  
\usepackage{mathptmx}                  
\usepackage{times}                     
\usepackage{cite}                      
\usepackage{tabu}                      
\usepackage{booktabs}                  

\usepackage{listings}
\usepackage{supertabular}
\usepackage{enumitem}
\usepackage{multirow}
\usepackage{tabularx}
\usepackage{xcolor}
\usepackage{xspace}
\usepackage{amsmath}
\usepackage{color}
\usepackage{soul}

\colorlet{punct}{red!60!black}
\definecolor{background}{HTML}{EEEEEE}
\definecolor{delim}{RGB}{20,105,176}
\colorlet{numb}{magenta!60!black}

\lstdefinelanguage{json}{
    basicstyle=\normalfont\ttfamily,
    numbers=left,
    numberstyle=\scriptsize,
    stepnumber=1,
    numbersep=8pt,
    showstringspaces=false,
    breaklines=true,
    frame=lines,
    backgroundcolor=\color{background},
    literate=
     *{0}{{{\color{numb}0}}}{1}
      {1}{{{\color{numb}1}}}{1}
      {2}{{{\color{numb}2}}}{1}
      {3}{{{\color{numb}3}}}{1}
      {4}{{{\color{numb}4}}}{1}
      {5}{{{\color{numb}5}}}{1}
      {6}{{{\color{numb}6}}}{1}
      {7}{{{\color{numb}7}}}{1}
      {8}{{{\color{numb}8}}}{1}
      {9}{{{\color{numb}9}}}{1}
      {:}{{{\color{punct}{:}}}}{1}
      {,}{{{\color{punct}{,}}}}{1}
      {\{}{{{\color{delim}{\{}}}}{1}
      {\}}{{{\color{delim}{\}}}}}{1}
      {[}{{{\color{delim}{[}}}}{1}
      {]}{{{\color{delim}{]}}}}{1},
}

\usepackage{fontawesome}

%% file: commands.tex
\newcommand{\revise}[1]{#1}

\newcommand{\PP}[1]{
\vspace{5px}
\noindent{\bf\textsc{#1}}\xspace
}

%% file: variables.tex
\newcommand{\app}{SafetyLens\xspace}
\newcommand{\fusa}{Functional Safety\xspace}
\newcommand{\fusashort}{functional safety\xspace}

\newcommand{\projecta}{Project-A\xspace}
\newcommand{\projectb}{Project-B\xspace}
\newcommand{\projectc}{Project-C\xspace}

\newcommand{\users}{domain experts\xspace}
\newcommand{\user}{domain expert\xspace}

\newcommand{\usera}{Chris\xspace}
\newcommand{\userb}{Kyle\xspace}
\newcommand{\userc}{Parker\xspace}

\definecolor{orange}{RGB}{237,125,49}
\definecolor{purple}{RGB}{112,47,160}
\definecolor{pink}{RGB}{245,114,255}
\definecolor{black}{RGB}{0,0,0}

%% file: src/sections/introduction.tex
Modern automobiles come integrated with a complex suite of systems (e.g. Anti-lock Braking System) that deploy hardware components integrated with millions of lines of software code. In a system that is not designed properly, the sheer number and complexity of components can result in failures posing risks which if not appropriately handled can be detrimental to the safety of the vehicle, its occupants, and the surrounding environment.
Recalls can tarnish a manufacturers brand,  straining customer trust and loyalty. It is important to design and manufacture vehicles in a way that may assist in reducing the risk. 
To analyze a given design, several systems engineering analysis techniques exist, which include but are not limited to Failure Mode Effect Analysis\cite{fmea-definition}, Fault Tree Analysis\cite{fta-definition}, etc.

Our work here is focused on \fusa in the automotive sector. To understand what \fusa is and how domain experts use it to ensure vehicle safety, consider an example scenario of the \textbf{``Adaptive Cruise Control''} system whose objective is to accelerate (or decelerate) the vehicle to maintain a reference speed as set by the driver. However, an improperly designed system could malfunction causing the vehicle to continuously accelerate (or decelerate). For example, when the vehicle is \textbf{overtaking}, if acceleration was to continue, and not be controlled by the driver, it can lead to a collision. To minimize risk associated with this mishap, a goal to \textbf{not accelerate more than required} is set. To achieve this goal, a requirement to be able to \textbf{override acceleration when brake pedal is pressed} is defined which in turn requires \textbf{monitoring of the brake pedal sensor}. Furthermore, this may require software code to \textbf{compare brake pedal sensor outputs for confidence} and possibly \textbf{additional sensors}. The process of setting these goals, testing the conditions, and analyzing the results are all part of \fusa.

\begin{figure*}[t]
    \centering
    \includegraphics[width=\textwidth]{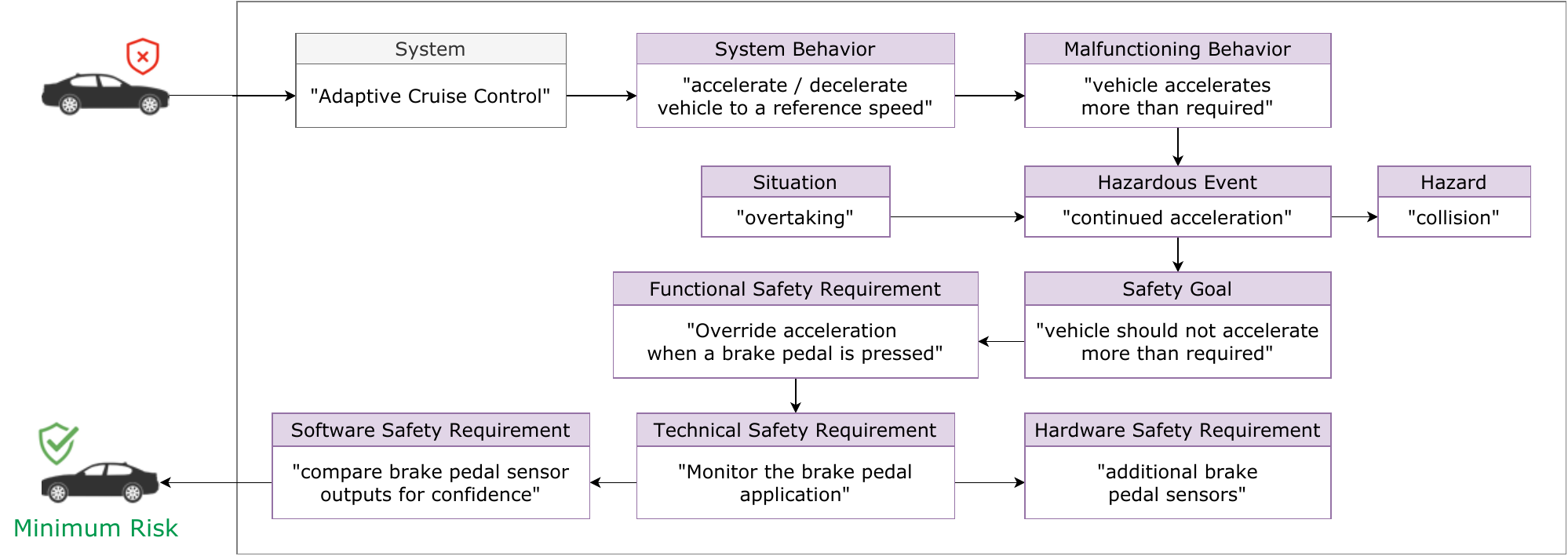}
    \caption{A hypothetical scenario of an \emph{``Adaptive Cruise Control''} system on which \fusa is performed. \textcolor{purple}{Purple} boxes indicate the \emph{Element} \emph{Types} described in Table~\ref{table:element-type-table}.}
    \label{fig:fusa-summary-and-data}
    \vspace{-1.5em}
\end{figure*}

The above example showcases one scenario of the \emph{``Adaptive Cruise Control''} system. There are other vehicle systems (e.g., Anti-lock Braking System, Parking Assist System) with their potential scenarios (similar or different sets of expected functions, associated risks, and safety goals). Finally, car companies have multiple car models, with updates from year to year, each with potentially different systems.
This makes \fusa datasets large. Figure~\ref{fig:fusa-summary-and-data} illustrates the \emph{``Adaptive Cruise Control''} scenario in the form of a network with each node representing an entity. The \textcolor{purple}{purple} boxes indicate the corresponding \fusa terminologies for the entities in the example (described in Section~\ref{fusa_concepts}).

\fusa domain experts are engineers and analysts who can identify and assess hazards, set goals to minimize risks associated with these hazards, and define safety requirements to achieve these goals. They must also ensure smooth implementation of these goals by asking questions such as \emph{``Are there any scenarios where there is an unreasonable risk for which there is no safety goal?''}, \emph{``Are there hazards in the current project that were also addressed in another project? If so, what safety goals were identified for them?''}, and others.

We designed and developed \app, a visual data analysis tool to assist \users to better understand \fusashort datasets. Through a three-month-long user-centered design process comprising interviews, demos, and discussions with \users from a multinational automobile company, we derived design goals and user tasks that drove our work. After a final phase of user feedback on the prototype, \app is currently being deployed.

The primary contributions of this paper include (i) design goals and challenges in supporting visual analysis of \fusa datasets, (ii) \app, a visual data analysis tool that helps \users visualize and interact with \fusashort datasets, (iii) usage scenarios that illustrate how \app can aid the current analysis processes of \users, and (iv) feedback from \users about the design, functionality, and impact of \app.

%% file: src/sections/industrial_background.tex


\fusa analyses are performed in several domains such as Military Aviation \cite{iec61508}, Space \cite{iso12207}, Medical \cite{iec62304}, Automotive \cite{iso26262}, etc. Our work here is focused on \fusa of vehicles within the automotive systems engineering domain. ISO 26262 titled ``Functional Safety - Road Vehicles''\cite{iso26262} is the international standard for the design and development of automotive electrical and electronic systems that makes Functional Safety a part of the automotive product development life-cycle. It is a risk-based safety standard, where the risk from hazardous situations is qualitatively assessed leading to the definition of safety measures that minimize these risks.

\subsection{Concepts}
\label{fusa_concepts}
This section introduces concepts and terminology associated with \fusa from the perspective of ISO 26262. These also describe the \fusa datasets used in \app. When modeled as a network, they make up the nodes, links, and attributes.

\PP{System} refers to a classification of automotive electrical and electronic systems such as ``Adaptive Cruise Control'' and ``Chassis'' on which \fusa is performed.

\PP{Project} refers to an instance of a \textit{System}, e.g., \emph{``Chassis for Model Z''}. This is also an instance of a \fusa dataset.

\PP{Element} refers to an entity within a \textit{Project} (e.g., \emph{``Vehicle should not accelerate more than required.''}). Elements constitute nodes in a \fusa network.

\PP{Type} refers to a group of similar \textit{Elements} within a \textit{Project}, e.g. \emph{Malfunctioning Behaviors (MB)}, described in Table~\ref{table:element-type-table}. Type is an attribute of the nodes in \fusa networks.

\PP{Relation} refers to the relationship between two \emph{Elements}. For example, a \emph{Relation} \emph{``associatedSG''} indicates that a \emph{Safety Goal (SG)} has been assigned to a \emph{Hazardous Event (HzE)}. Relations constitute the links in \fusa networks. Some of these Relations are shown in the Links column in Figure~\ref{fig:summary-view}.

\PP{ASIL}
{Automotive Safety Integrity Level}, or ASIL, is a nominal attribute assigned to an \emph{Element} (node) indicating its relative risk level. It can take one of four values: \{A, B, C, D\}. An \emph{Element} that is assigned an \emph{ASIL=D} indicates the highest risk. \emph{ASIL=A} indicates the lowest risk. A special value \emph{QM} indicates that quality management processes are sufficient to manage the identified risk \cite{iso26262}.

\input{"src/tables/node_types.tex"}

\subsection{Workflow}
\input{"src/sections/workflow.tex"}

\subsection{Data}
Each \textit{Project} constitutes an instance of a \fusa dataset. As seen in Figure~\ref{fig:fusa-summary-and-data}, this can be modeled as a network with nodes and links. Each node represents a \fusashort \textit{Element} with seven attributes: \textit{\{ID, Name, Type, ASIL, Severity, Exposure, Controllability\}}. A link between two nodes represents the \textit{Relation} between two \textit{Elements} and has three attributes: \textit{\{Source, Target, Relation\}}.

The dataset has several properties. First, a node of a particular \emph{Type} can be connected to zero (orphan), one, or more nodes of some other \emph{Type}. Second, \emph{Functional Safety Requirement (FSR)} and \emph{Technical Safety Requirement (TSR)} nodes can be connected to nodes of the same \emph{Type} as themselves. Third, a dataset can comprise nodes and links that are common to other projects (e.g., the same hazard (\emph{HzE}) can arise due to different malfunctioning system behaviors (\emph{MB})). Fourth, these common nodes may be assigned different ASIL values.
%
Finally, a standard \fusa network may consist of multiple nodes and links with the vehicle itself having multiple systems. This scale and complexity makes \fusa datasets 
large, and make the tasks of our \users challenging. 
%

%% file: src/tables/node_types.tex
\begin{table}[t]
    \centering
    \begin{tabularx}{\columnwidth}{p{0.25\columnwidth} | l | X }
        \textbf{Element Type} & \textbf{Abb.} & \textbf{Description} \\ 
        \hline
        System Behavior & SB & An expected behavior of an electrical/electronic \emph{System} in a vehicle.  \\
        Malfunctioning Behavior & MB & A failure or unintended behaviour of an item with respect to its design intent. \\
        Situation & Sit & A vehicular operational scenario.  \\
        Hazard & Haz & A potential source of harm.  \\
        Hazardous Event & HzE & A dangerous condition arising out of a \textit{Situation} and a \textit{Hazard}.  \\
        Safety Goal & SG & A requirement to eliminate (or minimize) \textit{Hazards}.  \\
        Functional Safety Requirement & FSR & A requirement to achieve the functional or behavioral aspects of a \textit{Safety Goal}.  \\
        Technical Safety Requirement & TSR & A technical requirement to meet a \textit{Functional Safety Requirement}.  \\
        Software Safety Requirement & SSR & A software-level requirement.  \\
        Hardware Safety Requirement & HSR & A hardware-level requirement.  \\
    \end{tabularx}
    \caption{A description of the \emph{Types} of \emph{Elements} in the context of automotive \fusa. Other Element \emph{Types} such as \emph{Physical}, \emph{Logical}, and \emph{Assumptions} exist but are beyond the scope of this paper.}
    \label{table:element-type-table}
    \vspace{-1.5em}
\end{table}

%% file: src/sections/workflow.tex
\fusa is a broad process ranging from the design through to the manufacturing and quality management of vehicles. The processes relevant to our work here comprise the following:

\PP{Hazard Analysis and Risk Assessment (HARA): }
Domain experts first determine potential hazards that can be triggered by the malfunctioning of one or more system components and/or processes. These hazards are classified based on their \emph{Severity (S)}, \emph{Exposure (E)}, and \emph{Controllability (C)}. Severity is an estimate of the extent of harm to one or more individuals that can occur in a potentially hazardous situation. Exposure is a measure of the probability of being in a hazardous situation. Controllability is an estimate of the driver's ability to avoid harm or damage through timely reactions. These are qualitative ratings with Severity=\emph{\{S1, S2, S3\}}, Exposure=\emph{\{E0, E1, E2, E3, E4\}}, and Controllability=\emph{\{C0, C1, C2, C3\}}. Based on these, an \emph{Automotive Safety Integrity Level (ASIL)} is calculated and assigned to the corresponding \emph{Hazardous Event (HzE)}. To minimize the risk associated with this hazard, a \emph{Safety Goal (SG)} is formulated.

\PP{Functional Safety Concept (FSC): }
From the formulated \emph{SG}, \emph{Functional Safety Requirements (FSR)} are derived.
These define a system architecture to achieve the \emph{SGs}.

\PP{Technical Safety Concept (TSC): }
The technical safety concept comprises all \emph{Technical Safety Requirements (TSRs)} along with their allocations to system, hardware or software elements. These are refinements of the corresponding \emph{FSRs}.
  
In practice, \fusa analysts and engineers can use a combination of commercial and open source tools to perform their day-to-day tasks. The \fusa processes can either be performed and documented in such tools or using traditional approaches such as Excel sheets and Word documents. For instance, our domain collaborators have utilized a domain-specific language (DSL) for ISO 26262 that enables them to import artifacts from HARA, FSC, TSC into a model. They use this model to generate Word documents with information on \fusashort that was performed on a system.

There can be several challenges in the above workflow, especially in the translation of an assigned ASIL into a Technical Safety Concept. Assigning an ASIL is a complex process that can involve discussions among multiple teams. Further, ASIL assignments can be modified at different \fusashort stages and points in time. 
The introduction of model-based systems engineering (MBSE) approaches may solve part of the problem as collaborators can now exchange information using a common language like SysML\cite{ramos2012mbse}. However, this may still have limitations when the scale and complexity of data increases. Finally, analyzing multiple projects to detect shared components and problems is not supported, and hence, a core focus of \app.

%% file: src/sections/related_work.tex

\PP{Network Visualizations}
Since a \fusashort dataset can be modeled as a network, we explored existing literature in network visualization systems for inspiration. There are a number of open source, free, or commercially available software \cite{batagelj1998pajek,auber2004tulip,bastian2009gephi,borgatti2002ucinet,smoot2011cytoscape}, toolkits \cite{o2003jung,hagberg2008exploring}, and research prototypes \cite{shneiderman2006network,kang2007netlens,van2009search}. Multiple survey reports \cite{von2011visual,herman2000graph,beck2017taxonomy} have not only summarized the state-of-the-art of network visualizations and techniques but also discussed their general evolution. We review Saket et al.'s evaluation study that indicated node-link-group visualizations as more ``enjoyable'' than node-link visualizations \cite{saket2016comparing} as a potential technique to represent the \fusashort network. \revise{A core \fusashort task is to compare projects (networks) hence we review Gove, R.'s V3SPA, a visual analysis tool (for security policy workers) to explore and find differences between large complex networks (SELinux and SEAndroid security policies) \cite{gove2016v3spa}. Another core \fusashort task is to find and visualize connections and/or paths between elements (nodes) hence we also review existing work in route tracing techniques. Candela et al. have developed Radian, an internet probe that helps visualize trace route paths \cite{candela2018radiantraceroutes}. Fischer et al. have developed Vistracer to investigate routing anomalies in traceroutes \cite{fischer2012vistracer}. Zhao et al have developed MissBin which infers the existence of unseen (missing) links based on currently observed ones by involving the user to sensemake the predicted results \cite{zhao2019missbin}.}

\PP{Set Visualizations}
Another \fusashort task involves comparing graphs to find common nodes and links. For this, we explored existing literature in set visualization. Euler and Venn diagrams are the most common methods to visualize sets and their intersections. Euler diagrams represent each set as a geometric shape and show the intersections by overlapping the shapes. Venn diagrams are like Euler diagrams but show all intersections, including empty ones. Sadana et al. \cite{sadana2014onset} developed Onset to represent large-scale binary set data. Lex et al.'s UpSet technique used a \emph{set view} and \emph{element view} to visualize intersections in a matrix layout introducing aggregates based on groupings and queries\cite{lex2014upset}. Alsallakh et al.'s survey paper discusses the state-of-the-art of set visualization systems and techniques\cite{alsallakh2016state}.

\PP{Visual Analytics in the Automotive Domain}
Perhaps most relevant to this paper is previous work on visualization applied to the automotive industry. These had been mostly in the context of scientific visualization, computer aided design (CAD), and virtual reality\cite{stevens2007visualization}. Considerably less work was dedicated to the systems engineering domain until recently. Basole et al. developed visual analytics tools to help users perform Complex Engineered System (CES) design analysis tasks helping stakeholders with visualizations of complex design models \cite{basole2015visual} and understand Failure Mode Effect Analysis (FMEA) data by modelling and visualizing it as a network \cite{basole-datahawk}. Sedlmair et al. have developed multiple tools leveraging visual analytics in the electronic engineering domain for vehicle development and testing. One such tool is Cardiogram, that helps engineers debug millions of recorded messages from safety-critical in-car communication networks and ensure that they are error-free \cite{sedlmair2011cardiogram}. RelEx helps engineers specify and optimize traffic patterns for in-car communication networks\cite{sedlmair2012relex}. MostVis facilitates exploration of MOST function catalogs which was otherwise infeasible using paper and existing database interfaces\cite{sedlmairr2009mostvis}. Another tool was a dual-view visualization system that helped diagnostics of in-car communication networks \cite{sedlmair2008dualview-viz}. 

\PP{\fusa} is a critical operation in automotive systems engineering whose methodologies\cite{ismail2013research,burton2012automotive,iso26262} and applications \cite{chang2014assessing,hommes2012review,birch2013safety,ward2013threat} have been extensively researched in literature. 
Several processes in systems engineering such as design reviews are often still conducted using unstructured documents (e.g., Excel sheets \cite{msexcel}) which is non-standard, inconsistent, and can cause miscommunication \cite{fischer1998mbse-automobiles}. Part of the problem was solved by the introduction of model-based systems engineering (MBSE) approaches that encourage information exchange using a common language \cite{ramos2012mbse} (e.g., SysML\cite{sysml-book}). There exist several tools now, both commercial \cite{sysml-medini,sysml-nomagic} as well as open-source \cite{sysml-papyrus,sysml-modelio} that support commonly performed tasks in systems engineering. 
However, these advances have resulted in more complex models and user tasks. 

\revise{In general, these existing tools focus their user experience and functionality on the authoring of the information. These are valuable for creating the information, but lack user interface affordances for communicating and exploring the data for various tasks described in subsequent sections. Moreover, these applications cannot be customized to support newer or more advanced types of tasks. For example, comparing multiple projects or visualizing the ASIL trace and decomposition cannot be easily accomplished in these tools. The potential application of viewing \fusashort data in a visual analysis tool is under-explored, motivating the need for this work.}

%% file: src/sections/domain_exploration/main.tex
\begin{figure}[t]
    \centering
    \includegraphics[width=\columnwidth]{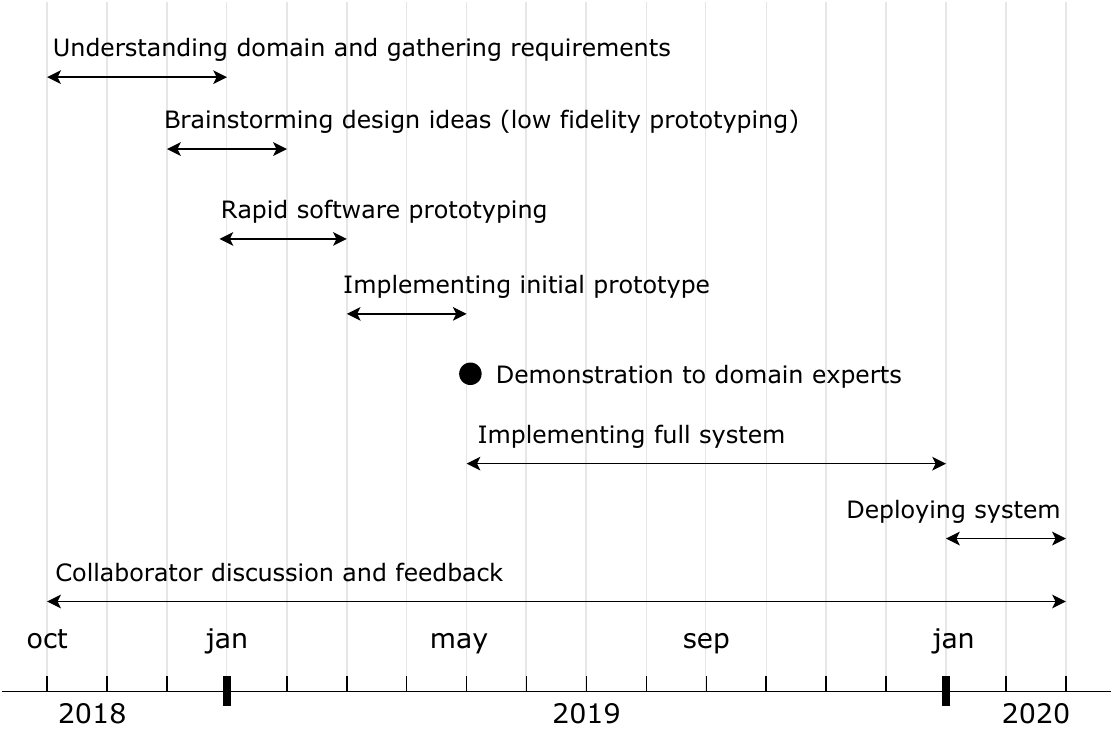}
    \caption{The high-level phases and milestones of the design process.}
    \label{fig:research-timeline}
    \vspace{-1.5em}
\end{figure}

We adopted a user-centered design methodology and conducted a series of design activities to learn about \fusa, the requirements and tasks of the \users, and distill design goals that ultimately ground \app (see Figure~\ref{fig:research-timeline}). This process included formative evaluation, design sessions, and iterative prototype development.  

First, we conducted interviews with \fusashort engineers, analysts, and supervisors to understand the domain requirements and user tasks. We conducted these interviews remotely (using teleconference applications with video and screensharing capabilities) over the course of three months, with one session per week. Generally, the sessions lasted 1 hour, with 1-4 domain experts on the calls. These sessions consisted of semi-structured interviews that started with questions to learn about their tasks, roles, and responsibilities. We gained an understanding of the current design ecosystem of tools, tasks, and datasets. We learned about the questions they have to answer as part of their daily activities, and specifically how those relate to the different datasets, databases, and other sources of information (e.g., manuals, regulations, etc.). We learned about the current workflow and identified 25 commonly performed tasks. 

Second, we began design exercises to explore potential visualization and interaction techniques. We sketched ideas on paper and shared them with our collaborators during subsequent sessions. These sketches included potential visualization techniques (e.g., tree maps, force directed networks, hive plots), UI widgets (e.g., sliders, dropdowns, buttons), UI layouts (e.g., panels, grids), workflows, and interactions. These were low-fidelity designs sketched quickly but detailed enough to catch errors and slips that could surface later. During these sessions, we collaboratively brainstormed on the pros and cons of each design which eventually resulted in multiple changes and refinements. We also realized these designs by developing rapid software prototypes with a dual purpose of exploring potential technologies for the tool (such as software libraries and packages) and evaluating the feasibility of the designs. We deployed these prototypes for the domain experts to interact and evaluate them based on their usability and visual look and feel, with small subsets of their data integrated into the prototypes. This helped discard less-useful designs, fine-tune our initial set of tasks, and derive 6 design goals.

Third, we designed and developed an initial prototype of \app and conducted an in-person demo and feedback session. This consisted of two half-day sessions with \users \revise{comprising functional safety analysts, project managers, and personnel from application-specific teams (e.g., chassis, power-train) who were} split into two groups. The first session included 15 participants (7 in-person, 8 online), and the second one included 20 participants (6 in-person, 14 online). Each session consisted of a presentation of the design goals and supported user tasks, followed by a demonstration of the prototype. One researcher presented the prototype and led the discussion, while a second took notes on participants' feedback. The participants engaged in thoughtful discussions about the demonstrated usage scenarios. They had access to sticky notes to keep track of comments throughout the presentation (with remote participants being able to use chat and other messaging services to give their feedback). Also, screen-shots were given to each participant to annotate with comments, changes, or other visual cues about how to improve the interface and interaction design.

Finally, a whiteboard was used by the in-person participants to illustrate how attributes from different databases can be integrated into \app. These discussions led to refinements to existing features, as well as new tasks that the tool could support. The user tasks and design goals presented below are those that resulted from this second round of user feedback.

\subsection{User Tasks}
\input{"src/sections/domain_exploration/user_tasks.tex"}

\subsection{Design Goals}
\input{"src/sections/domain_exploration/design_goals.tex"}

%% file: src/sections/domain_exploration/user_tasks.tex
User tasks which \app should support are:
\begin{itemize}[noitemsep,topsep=0pt]
    \item Discover patterns within projects; e.g., a project has elements that are mostly assigned an ASIL=D (high risk).
    \item Find missing (otherwise \emph{must-have}) links between nodes; e.g., \{\emph{MB} $\rightarrow$ \emph{HzE}\}, \{\emph{HzE} $\rightarrow$ \emph{SG}\}, \{\emph{SG} $\rightarrow$ \emph{FSR}\}, \{\emph{FSR} $\rightarrow $ \emph{TSR}\}.
    \item Look up and analyze a node's end-to-end traceability (i.e., the extent to which \fusashort elements have been defined); e.g. \{\emph{MB} $\rightarrow$ \emph{HzE} $\rightarrow$ \emph{SG} $\rightarrow$ \emph{FSR} $\rightarrow$ \emph{TSR}\}.
    \item Compare ASILs by analyzing their decomposition into their \emph{S-E-C} (Severity, Exposure, and Controllability) constituents.
    \item Find common nodes and links among projects; (e.g., \emph{Hazards (Haz)} that exist in multiple projects).
    \item Discover anomalies across projects; (e.g., the same \emph{Hazardous Event (HzE)} is assigned a different ASIL across projects).
    \item Compare key metrics across projects; (e.g., counts and distribution of nodes and links within projects).
    \item Get an overview of current and past projects to monitor status.
\end{itemize}

%% file: src/sections/domain_exploration/design_goals.tex
From the design exercises and requirements gathering activities described above we identified the following design goals.

\PP{DG1: Facilitate Exploration of a Project }
User tasks such as ``Find missing links'', ``Lookup and analyze a node’s end-to-end traceability'' are network exploration tasks. Hence, we modeled \fusa data as a network and derived this design goal to support tasks like fetching node details on hover, finding adjacent nodes, and finding paths from one node to another based on the taxonomies by Lee et al. \cite{lee2006tasktaxonomy} and Pretorius et al. \cite{pretorius2014tasks}.

\PP{DG2: Facilitate Comparison among Projects }
Teams within \fusashort can be system-specific and may not always be aware of the day to day progress made by other teams. This may result in duplicate work (e.g., a team may re-implement an artifact from scratch instead of re-using the one already implemented by another team, or for another vehicle). A core goal for \app, thus, was to provide a unified interface where users can explore and compare multiple projects to find shared (common) nodes, links, or even sub-graphs (combination of nodes and links). This unified interface can foster better collaboration among teams while also saving time and resources for the organization.

\PP{DG3: Discover Patterns and Anomalies }
The \users we spoke to make use of existing commercial and open source tools. These tools support basic exploration and comparison of \fusa projects but may fall short when the scale and complexity of data increases. Thus, \app should support discovering interesting patterns within and across projects. 

\PP{DG4: Facilitate Traceability and Decomposition of ASILs }
An important task for \users is to trace the ASIL from one node to another (e.g., \{\emph{MB} $\rightarrow$ \emph{HzE} $\rightarrow$ \emph{SG}  $\rightarrow$ \emph{FSR} $\rightarrow$ \emph{TSR}\}). Since ASILs determine the extent of safety mechanisms for elements, any discrepancy such as an element assigned an ASIL=A instead of ASIL=D is an important concern. To diagnose the problem, users should be able to decompose the ASIL into its \emph{Severity (S), Exposure (E), and Controllability (C)}. Thus, it is an important design goal for \app to unify tasks allowing users to efficiently detect, diagnose, and fix ASIL assignment issues.

\PP{DG5: Support different User Groups }
\fusa consists of engineers, analysts, and managers all of whom perform analysis and decision-making tasks at different levels. While engineers and analysts are concerned with lower element-level tasks within projects, managers are concerned with higher project-level tasks. \app should support both along the \fusashort organization hierarchy.

\PP{DG6: Provide a Summary of Key Metrics }
The user tasks suggested that key metrics should be readily available (e.g., total number of nodes, total number of nodes with ASIL=D, etc). 


In addition to these design goals, we had to consider other factors. Since \fusa datasets can get large, we had to design for scalability and performance. We utilized a graph database to persist the data and execute queries to offload computation on the browser. Further, we had to ensure our tool is easy-to-deploy and integrates well with existing systems and workflows (e.g., the use of linked spreadsheets that are local to groups and users as well as shared databases). Finally, our users are from the automotive domain and not visualization practitioners, hence \app needs a simple user experience.

\subsection{Design Considerations}
Since we modeled \fusa data as a network we considered several network visualization techniques. We implemented rapid prototypes of standard visualization techniques such as hive-plots, \revise{dot-matrix plots}, and parallel coordinate charts. These had a number of shortcomings. \revise{First, due to the number of links between nodes, edge-crossings can get messy. Second, positioning the nodes linearly in hive-plots made it difficult to discover patterns (e.g., distribution of ASILs). The dot-matrix plot enabled discovering patterns but give an impression that the \{x,y\} spatial coordinates of each node are of importance which was not the case. Third, parallel coordinates charts failed because the nodes had connections with more than two types of nodes.}

We implemented a force-directed network with a multi-body force algorithm generating node positions at random. However, similar nodes (e.g., \emph{Elements} with the same \emph{Type}) were scattered throughout the canvas making pattern discovery difficult. Inspired by Saket et al's study that found node-link-group diagrams more ``enjoyable'' than node-link diagrams \cite{saket2016comparing}, we augmented our visualization with circle packing (to group similar nodes together within a cluster), and collision detection (to prevent overlaps between clusters and nodes) algorithms to achieve our goal. To facilitate interactive exploration and ``fluidity''~\cite{elmqvist2011fluid}, we provide affordances to drag the clusters within the canvas, brush and link between projects by a lasso operation, and a control panel with search and filter direct manipulation widgets.

To support comparison between projects, we show multiple node-link-group diagrams aligned spatially. One idea was to have a single visualization with all projects on a shared canvas. However, this had a major shortcoming. Since projects comprise shared nodes with attributes that are specific to a project (e.g., Node A in Project A has an ASIL=B but ASIL=D in Project B), it was important to keep the nodes and hence the projects separate.
Our users are familiar with tabular representations of data (e.g., MS Excel). Keeping their visualization literacy in mind, we complemented each node-link-diagram with a data table. 
%
To facilitate comparison between projects, we considered juxtaposing the node-link-group diagram and the datatable within each project panel and vertically stacking the project panels themselves. Since comparing vertically is challenging, we flipped this approach and provided drag-and-drop affordances to reorder the project panels.

%% file: src/sections/system_overview.tex
This section describes the main views and functionality of \app. 

\subsection{User Interface}
 \app has three primary views: 1) Dashboard View, 2) Summary View, and 3) Compare View.
 
\begin{figure}[t]
    \centering
    \includegraphics[width=\columnwidth]{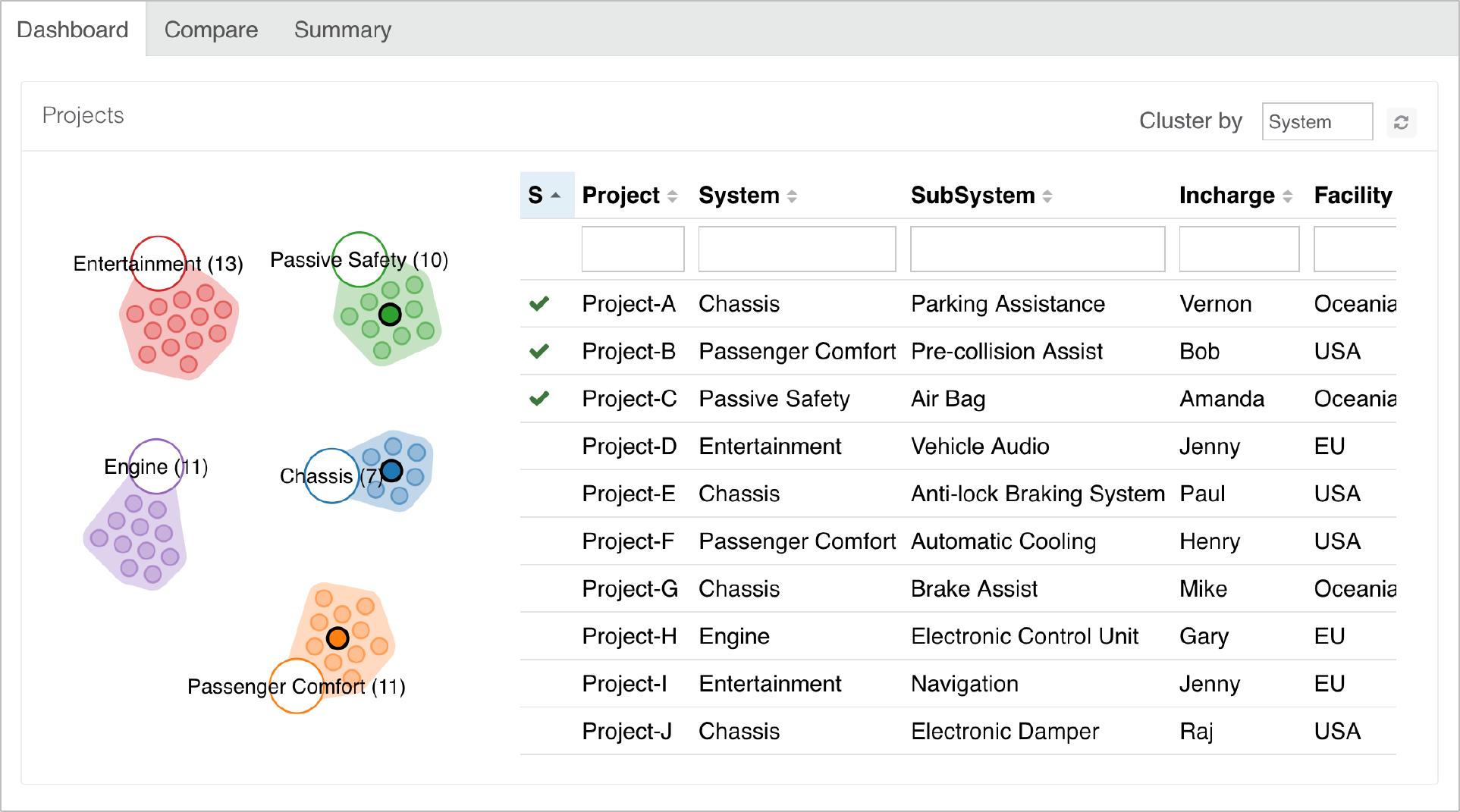}
    \caption{The \textbf{Dashboard View} shows available projects. \revise{The projects in the node-link group visualization on the left are clustered as well as colored based on their ``System'' (configured from the \textit{Cluster by} dropdown to the top-right of the Table View.}}
    \label{fig:dashboard-view}
    \vspace{-1.5em}
\end{figure}

\PP{Dashboard View}, shown in Figure~\ref{fig:dashboard-view}, is the landing page and provides an overview of the \fusa ecosystem. It consists of a clustered node visualization with each node representing a project. We complemented the visualization with a table to provide a familiar interface to users who are more comfortable with data in spreadsheets. The table consists of project-level information such as \{Name, Department, Project In-Charge, and Location\}. \app supports multiple starting points for users to analyze \fusashort data. 
By providing access to all projects within the organization, \app facilitates collaboration \textbf{(DG2)} among \users with different roles \textbf{(DG5)}.

\begin{figure}[t]
    \centering
    \includegraphics[width=\columnwidth]{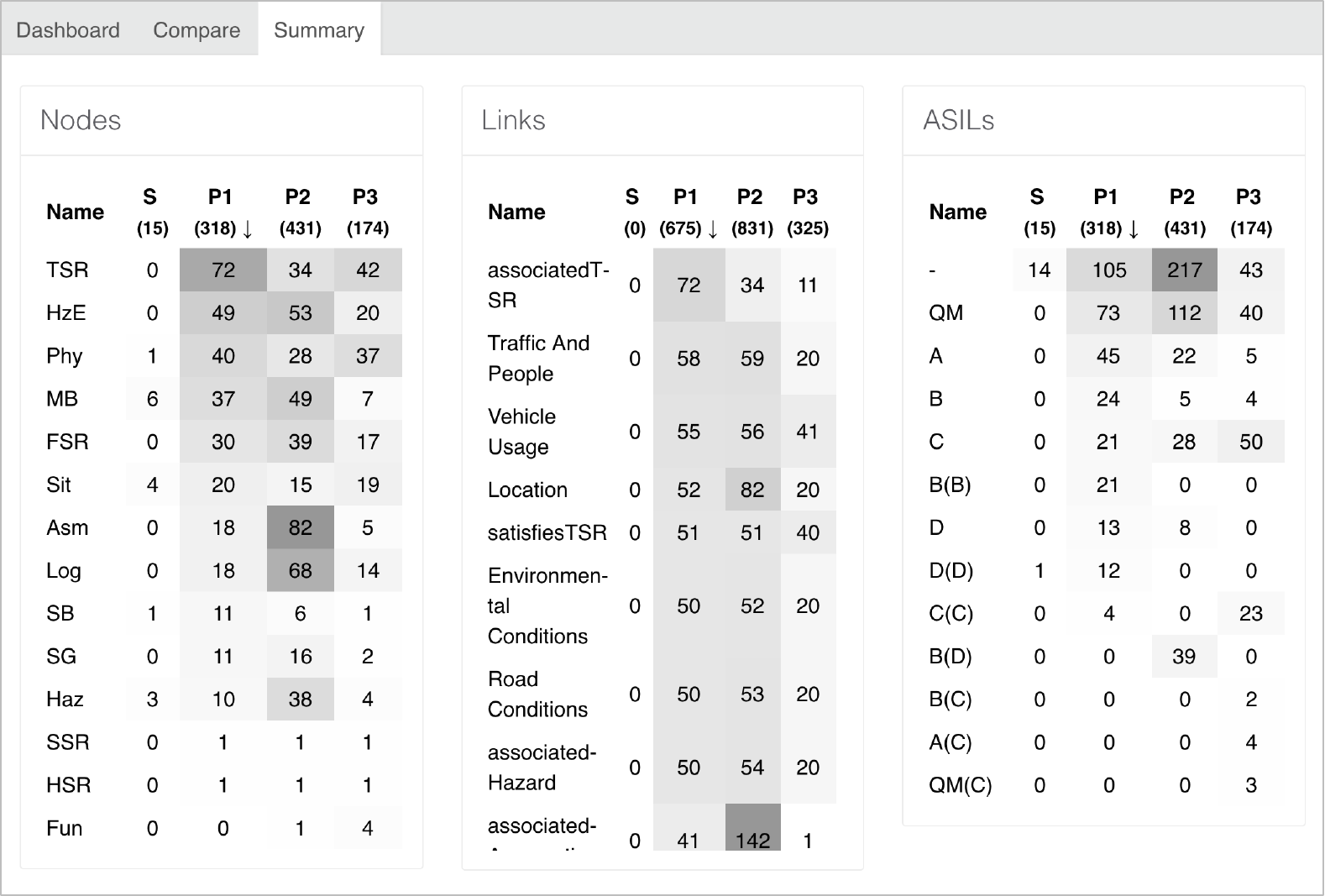}
    \caption{The \textbf{Summary View} shows \revise{three projects \textbf{P1}, \textbf{P2}, and \textbf{P3} and} the distribution of nodes by their \textit{Types}, links by their \textit{Relations}, and \textit{ASILs} by their values. The column \textbf{\textit{S}} is a count of shared entities (nodes and links) between the projects. \revise{Project \textbf{P1} has \textbf{318} nodes and \textbf{675} links. The projects share \textbf{15} nodes and \textbf{0} links among them.}}
    \label{fig:summary-view}
    \vspace{-1.5em}
\end{figure}

\PP{Summary View}, shown in Figure~\ref{fig:summary-view}, provides a summary of the projects selected in the Dashboard View \textbf{(DG6)}. There are juxtaposed heatmap visualizations for \textit{Node Type}, \textit{Link Relation},  and \textit{ASIL} respectively positioned next to each other. For each attribute table, projects are along the column axis and the corresponding attribute values are along the row axis. An additional column titled ``S'' is added to show the number of entities (nodes and links) that are shared among these projects. The cells show the per-project entity counts for the corresponding attribute, colored using a continuous color scale (\textit{white-to-gray}) to help the user discover patterns within as well as across projects \textbf{(DG3)}.

\PP{Compare View}, shown in Figure~\ref{fig:interface}, is the main view of the \app interface that allows users to simultaneously perform exploratory analysis of an individual project as well as a comparative analysis across multiple projects. It has four subviews (alphanumeric list enumerations match with those in Figure~\ref{fig:interface}):

\begin{enumerate}[label=(\Alph*),nosep,parsep=-5pt,topsep=-5pt,partopsep=-5pt]
    \item \PP{Project Panel}. Each imported project is rendered as a panel that shows the project title and utilities to (i) reset the panel state, (ii) export the panel as an image, (iii) toggle full-screen mode, and (iv) horizontally reorder the panel using drag and drop.
    
    \PP{Visualization canvas} shows a node-link group visualization. Each node (small circle) is an \emph{Element} of a \fusashort project. These nodes are clustered together based on attributes (e.g., \textit{Type}). The largest circles represent the group nodes and are labeled with the \emph{Type} and the number of nodes that are part of it. The boundary marking the extent of the groups (convex hull) is highlighted. The nodes are positioned in each others' vicinity using an implementation of the circle packing algorithm. The node size and color can be mapped to attributes such as \{ASIL, Type, Degree (number of edges to a node)\} which in the presence of multiple nodes across multiple projects will create visual clusters leading to discoveries of patterns and anomalies \textbf{(DG3)}. By default, the groups are positioned relative to each other based on the multi-body force and collision detection algorithms. However, these groups are fluid and can be re-positioned by dragging them around. Further, a control panel toggle button enables the users to position and fix these clusters to similar relative locations across all projects to aid comparison (e.g., the \emph{SB} cluster can be positioned to the top-right of the canvas across all projects).
    \revise{Section~\ref{interactions} describes all the interactions supported by this visualization.}

    \PP{Tab Layout}
    has three options: Nodes, Links, and Trace.
    \begin{enumerate}[label=(\arabic*), nosep,parsep=-5pt,topsep=5pt,partopsep=-5pt]
        \item \PP{Nodes Tab} shows a datatable with the nodes selected by the user with their \textit{\{Type, ASIL, Name, ID\}}.
        \item \PP{Links Tab} shows a datatable with the Links data with their \textit{\{Source, Relation and Target\}}.
        \item \PP{Trace Tab} (shown in Figure~\ref{fig:use-case-asil-trace}) allows the user to find and visualize if a path exists between two nodes as well as trace their ASILs \textbf{(DG4)} (e.g., tracing the ASIL from a \emph{System Behavior} to a \emph{Technical Software Requirement}). The user can set a node as \textit{Source} and another as \textit{Destination}. \app first checks if a path exists between the two nodes and overlays it onto the node-link group visualization. It also returns a linearized node-link diagram showing the entire route from source to destination. Below the node-link diagram is a heatmap showing the S-E-C (Severity-Exposure-Controllability) break up of the ASILs.
    \end{enumerate}

    Since the visualization canvas and the tab layout are vertically stacked within a project panel, we positioned each panel side by side. This way, \app would facilitate exploratory analyses within and comparative analyses across projects \textbf{(DG1, DG2)}.

    \item \PP{Shared View} shows the nodes and links that are common to / shared by all loaded projects \textbf{(DG2)}. These are computed by performing a set intersection operation across project graphs based on unique node identifiers. We show it in a separate view instead of overlaying or highlighting in the same view to make it easier for \users to begin their analysis.

    \item \PP{Details View} shows all attributes of a node as well as a table containing all nodes that are connected to this node.
    
    \item \PP{Control Panel} has three tabs: (i) Nodes, (ii) Links, and (iii) Config with various operations that can be performed (e.g., lookups, encodings, layout, preferences for tooltips, and more). 

\end{enumerate}


\revise{
\subsection{Interactions} \label{interactions}
The primary design goals for the user interactions are to facilitate brushing and linking between views while maintaining a simple and usable interface.
These interaction include:
\begin{enumerate}[label=(\Alph*),nosep,parsep=-5pt,topsep=-5pt, partopsep=-5pt]
    \item \PP{Hover} The visualization nodes and the datatable rows can be hovered to highlight neighbours as well as show more information in the Details View.
    \item \PP{Click} The visualization nodes and the datatable rows can be (left) clicked to toggle their selections. In addition, right-clicking on a visualization node opens a context-menu that supports operations such as \textit{Select}, \textit{Set as Source/Destination}, and others.
    \item \PP{Drag} The group nodes can be dragged to move the clusters within the visualization canvas. Similarly, the project panels themselves can be reordered to aid comparison.
    \item \PP{Selection} The visualization canvas can be used to draw free-form shapes via a lasso operation to select nodes and links.
    \item \PP{Search} The Search input field in the Control Panel can be used to ``lookup'' and ``select'' nodes by their names.
    \item \PP{Filter} The Control Panel has UI controls (e.g., radio buttons, checkboxes, dropdowns, interactive legends) that allow the user to ``filter'' and ``select'' the nodes and links in the visualization.
    \item \PP{Sort} The column headers of datatables in the Tab layout can be used to sort the selected nodes and links.
\end{enumerate}
}

\subsection{Implementation}
\revise{\app is implemented as a web application using Flask (a Python-based microframework)~\cite{pythonflask}, AngularJS~\cite{angularjs}, and D3.js~\cite{d3js} 
. The \fusa data is stored in OrientDB~\cite{orientdb}}. 

%% file: src/sections/use_cases.tex
We illustrate how \app can help \users visualize and interact with \fusa datasets. Consider three usage scenarios comprising several subtasks as performed by three hypothetical users - \usera (analyst), \userc (\usera's supervisor), and \userb (analyst) respectively. These usage scenarios were developed in collaboration with the domain experts from our interview studies to ensure domain relevance. Due to confidentiality concerns, project, node, and link names have been obfuscated. These scenarios are also illustrated in the accompanying video.

\subsection{General Exploratory Tasks}
\label{exploration}

    \usera and their colleagues have been actively working on \projecta (the ``Adaptive Cruise Control'' System) for a week. They have (i) identified several \emph{Malfunctioning Behaviors} and the potential \emph{Hazards} that could arise, (ii) assigned a risk level (ASIL) to these hazards and (iii) defined \emph{Safety Goals} and requirements (\emph{Functional Safety Requirements}) to minimize these risks. \usera wishes to perform a status check to determine 
    \revise{if the risks associated with the project components have been reduced to acceptable levels. This includes the following steps (or tasks)}. 
    
    \textbf{Find and address Orphan Nodes}
    \revise{It is important to ensure that the defined nodes are connected to other nodes (that is, are not orphan). By being aware of these orphan nodes, \usera can ensure that these are not implemented from scratch but instead reused from the same node's implementation in another project.}

\begin{figure}[t]
        \centering
        \includegraphics[width=\columnwidth]{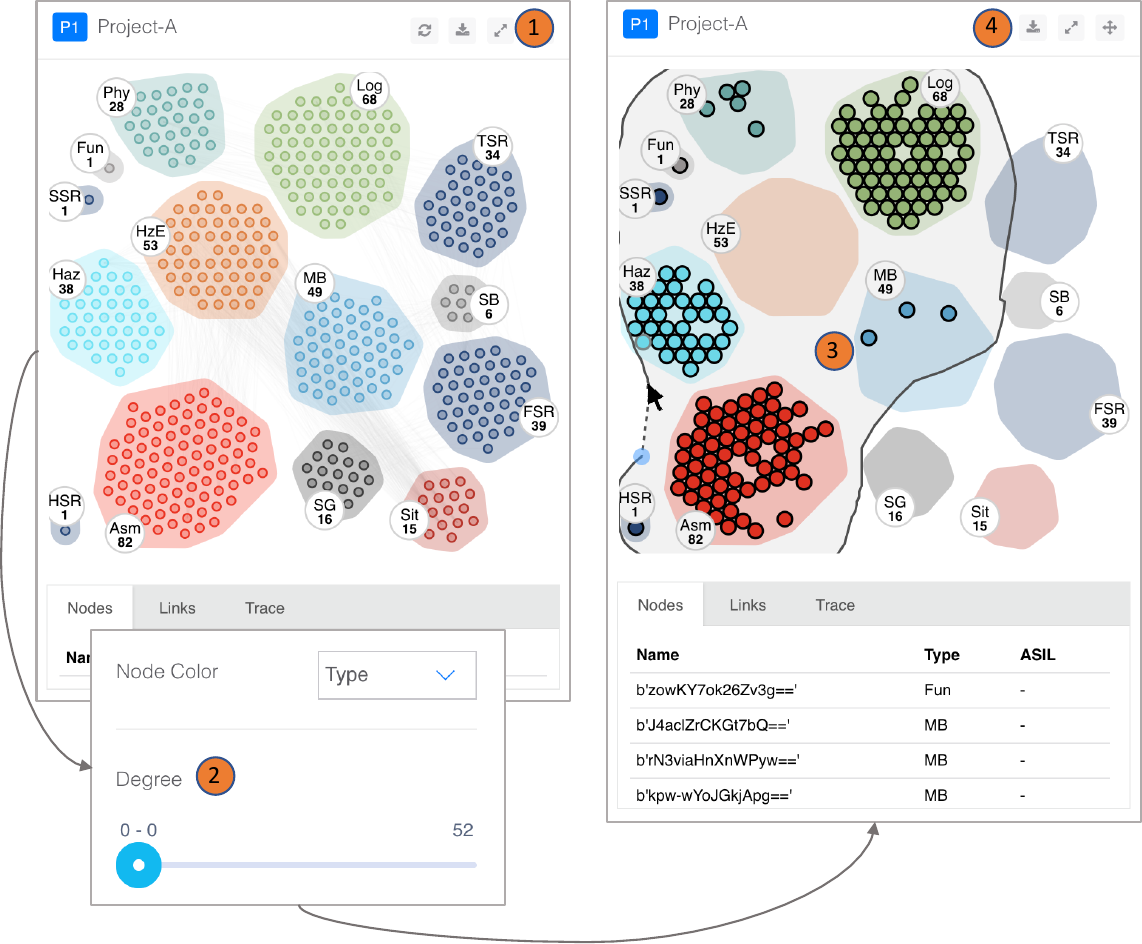}
        \caption{\textbf{Report Orphan Nodes and their Types}. \textcolor{orange}{\textbf{(1)}} Default State where nodes are colored by their \emph{Type}, \textcolor{orange}{\textbf{(2)}} Drag the degree slider to zero, \textcolor{orange}{\textbf{(3)}} Select the remaining nodes with a lasso operation; these are highlighted with a black stroke \revise{and also added to the nodes datatable below}, and \textcolor{orange}{\textbf{(4)}} Export the view as an image.}
        \label{fig:use-case-no-connections}
    \vspace{-1.5em}
    \end{figure}

    To accomplish this task, \usera first opens the Control Panel and drags the Degree slider to zero. This filters out the nodes that have at least one connection. Using a lasso operation, they select these nodes and export a snapshot of the visualization for future reference. Figure~\ref{fig:use-case-no-connections} summarizes this task.

   \textbf{Find and report nodes with an unassigned ASIL.}
   \revise{After addressing orphan nodes, \usera shifts their focus to assigning an ASIL to each node.}
   The ASIL determines the quality and quantity of safety measures that are undertaken to minimize associated risks. During this assignment process, it is possible that a few nodes are either missed or skipped for later. In both cases, the node is left without a valid ASIL.
    
\begin{figure}[t]
        \centering
        \includegraphics[width=\columnwidth]{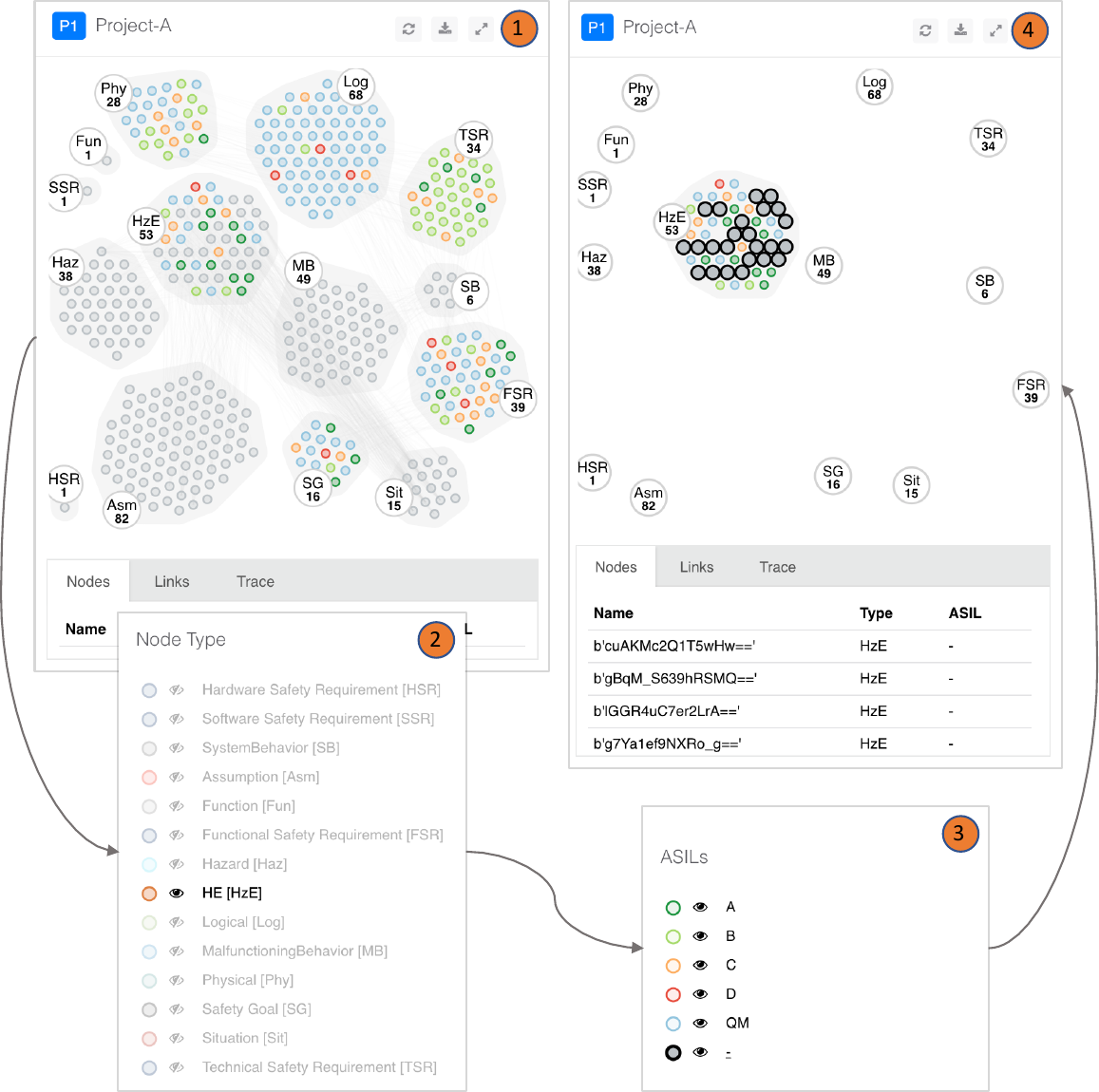}
        \caption{\textbf{Find elements that are not yet assigned an ASIL.} \textcolor{orange}{\textbf{(1)}} Default State where nodes are colored by their \emph{ASIL}, \textcolor{orange}{\textbf{(2)}} Deselect all Node Types except \emph{Hazardous Events (HzE)}, \textcolor{orange}{\textbf{(3)}} Select all nodes with unassigned ASILs, and \textcolor{orange}{\textbf{(4)}} The selected ASILs are highlighted with a \textbf{black} border.}
        \label{fig:use-case-asil-unassigned}
    \vspace{-1.5em}
    \end{figure}

    To verify this for \emph{Hazardous Events}, \usera opens the Control Panel and hides all node types except \emph{Hazardous Event}. From the ASIL legend, they select the nodes with an ASIL ``-'' (unassigned). Several nodes (gray) that do not have an ASIL are selected and added to the datatable below. \usera shares the selected nodes with their team to discuss and plan next steps. \usera's workflow is summarized in Figure~\ref{fig:use-case-asil-unassigned}.

    \textbf{Find Missing Links.}
    After analyzing \revise{orphan and ASIL-unassigned nodes}, \usera funnels their attention to analyze specific connections within the \fusa network to ensure that:
    \begin{enumerate}[nosep]
        \item Each \emph{Malfunctioning Behavior (MB)} should have a \emph{Hazardous Event (HzE)} identified for it.
        \item Each \emph{HzE} should have a \emph{Safety Goal (SG)} assigned to it.
        \item Each \emph{SG} should define a \emph{Functional Safety Requirement (FSR)}.
        \item Each \emph{FSR} should define a \emph{Technical Safety Requirement (TSR)}.
    \end{enumerate}

\begin{figure*}[t]
        \centering
        \includegraphics[width=\textwidth]{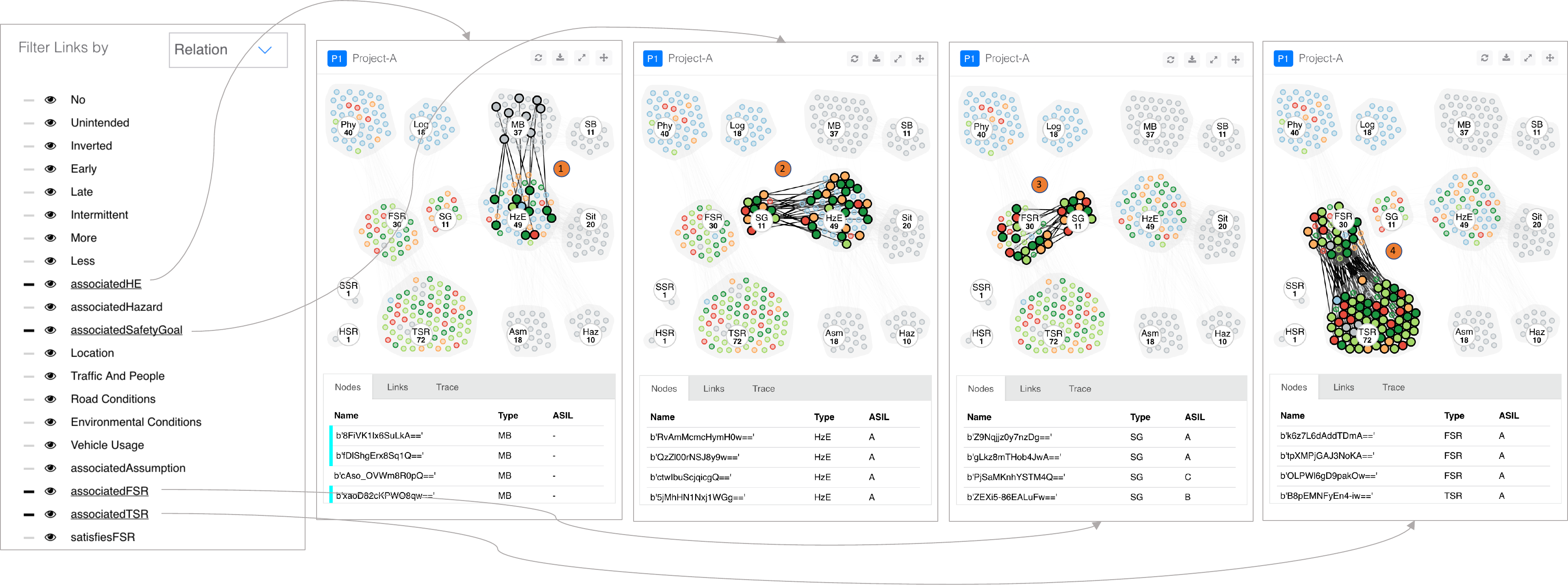}
        \caption{
            \textbf{Find Missing Links.}
            \textcolor{orange}{\textbf{(1)}} Each \emph{Malfunctioning Behavior (MB)} should have a \emph{Hazardous Event (HzE)}. 
            \textcolor{orange}{\textbf{(2)}} Each \emph{HzE} should have a \emph{Safety Goal (SG)}. 
            \textcolor{orange}{\textbf{(3)}} Each \emph{SG} should define a \emph{Functional Safety Requirement (FSR}. 
            \textcolor{orange}{\textbf{(4)}} Each \emph{FSR} should define a \emph{Technical Safety Requirement (TSR).}
            }
        \label{fig:use-case-link-relations}
    \end{figure*}

    A \user's role is to address such missing connections at the earliest. The links in the dataset comprise the \emph{Relation} attribute that determines the type of connection between two nodes, e.g. \emph{associatedHE} connects a \emph{Malfunctioning Behavior} with a \emph{Hazardous Event}. A unique Relation is defined for a connection between nodes of different Types. 
    To find these missing connections, \usera opens the Control Panel and switches to the Links tab. As illustrated in Figure~\ref{fig:use-case-link-relations}, they select the corresponding Relations from the legend that are highlighted in the visualization. \usera observes that:
    \begin{enumerate}[nosep]
        \item \textbf{\emph{associatedHE}}: \emph{HzEs} for more than half of \emph{MBs} are not identified.
        \item \textbf{\emph{associatedSafetyGoal}}: A few \emph{HzEs} do not have a \emph{SG} assigned.
        \item \textbf{\emph{associatedFSR}}: All \emph{SGs} have at least one \emph{FSR} defined.
        \item \textbf{\emph{associatedTSR}}: Very few \emph{FSRs} have no \emph{TSRs}.
    \end{enumerate}

    From the above usage scenario, \usera found several nodes without an ASIL, several nodes with zero as well as missing links. They conclude that even though their \projecta is under active development, there are gaps in their process that are hampering their efficiency. They take these results to their team to take required actions.
    
    \subsection{Trace ASIL}
    Another aspect of an analyst's role is to analyze the ASIL of a node in comparison with the ASILs of its connections (neighbor, neighbor's neighbor, and so on). Consider a \emph{Hazardous Event (HzE)} that is assigned the highest risk classification ASIL=D. It requires \emph{Safety Goals (SG)} that can sufficiently minimize the associated risk. By default, \emph{\{SGs\}} inherit the ASIL of the \emph{\{HzE\}} (\emph{\{FSRs\}} inherit the ASIL of the \emph{\{SG\}}, and \emph{\{TSRs\}} inherit the ASIL of the \emph{\{FSR\}}) but the analyst may choose to override it to achieve acceptable risk levels, making visual analysis of the ASIL trace an important task.
    
\begin{figure*}[t]
        \centering
        \includegraphics[width=\textwidth]{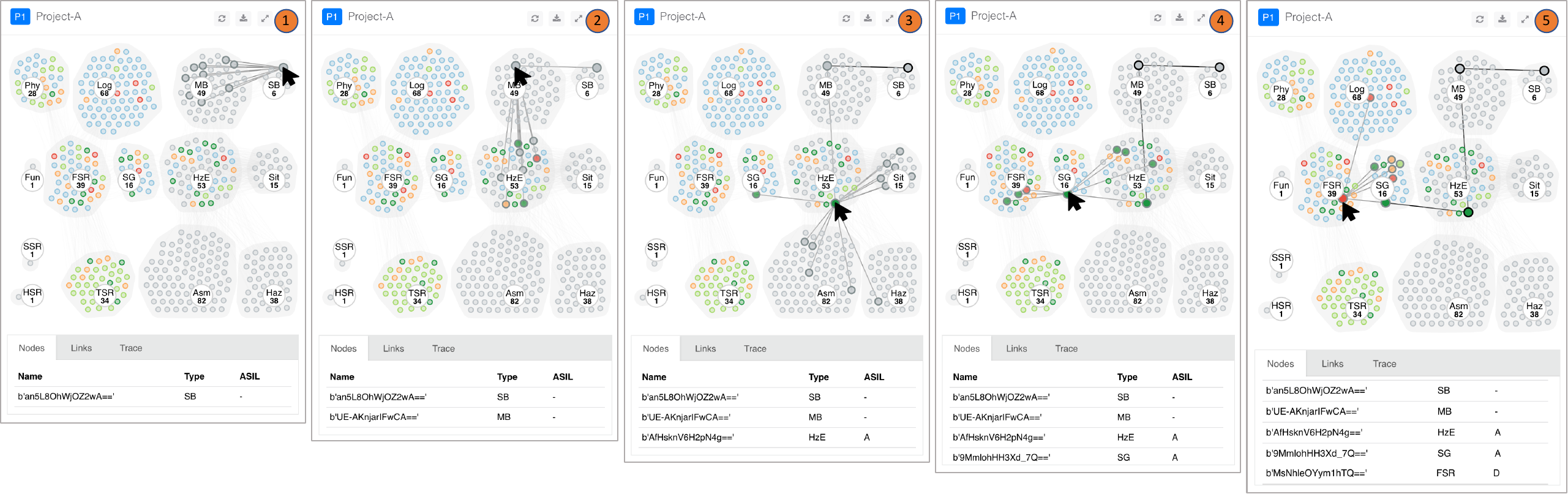}
        \caption{\textbf{Manually Trace paths.} \textcolor{orange}{\textbf{(1-5)}}: The user is manually tracing a path from a \emph{System Behavior (SB)} up to a \emph{Functional Safety Requirement (FSR)}. \textcolor{orange}{\textbf{(5)}}: The \emph{FSR} does not have a corresponding \emph{Technical Safety Requirement (TSR)}.}
        \label{fig:manual-traverse}
    \vspace{-1.5em}
    \end{figure*}

    \userb's role is to assess risks associated with \emph{HzEs} and qualitatively assign an ASIL to them. Their responsibilities also include analyzing the extent to which \fusa is achieved for a node, that is, finding paths between nodes that are not directly connected via a link. This way, a project's maturity (``Are there sufficient nodes and links implemented for this project?'') as well as completeness (``Most \emph{SBs} have a \emph{TSR} defined now; the project is nearing its completion.'') can be assessed. \userb performs the following tasks:
    
    \textbf{Manually Trace paths.}
    As illustrated in Figure~\ref{fig:manual-traverse}, \userb chooses a known \emph{SB} for further analysis. \textcolor{orange}{\textbf{(1)}} On hovering, they see several links connecting it with multiple \emph{MBs}. \textcolor{orange}{\textbf{(2)}} They select the \emph{SB} and hover on one of the \emph{MB} which highlights the links with \emph{HzEs}. \textcolor{orange}{\textbf{(3,4)}} They continue this process until they reach an \emph{FSR}. \textcolor{orange}{\textbf{(5)}} They find that the \emph{FSR} does not have a link with an \emph{TSR}. Since this is a requirement, this means \fusa is still incomplete for the \emph{SB} in (1).

\begin{figure}[t]
        \centering
        \includegraphics[width=\columnwidth]{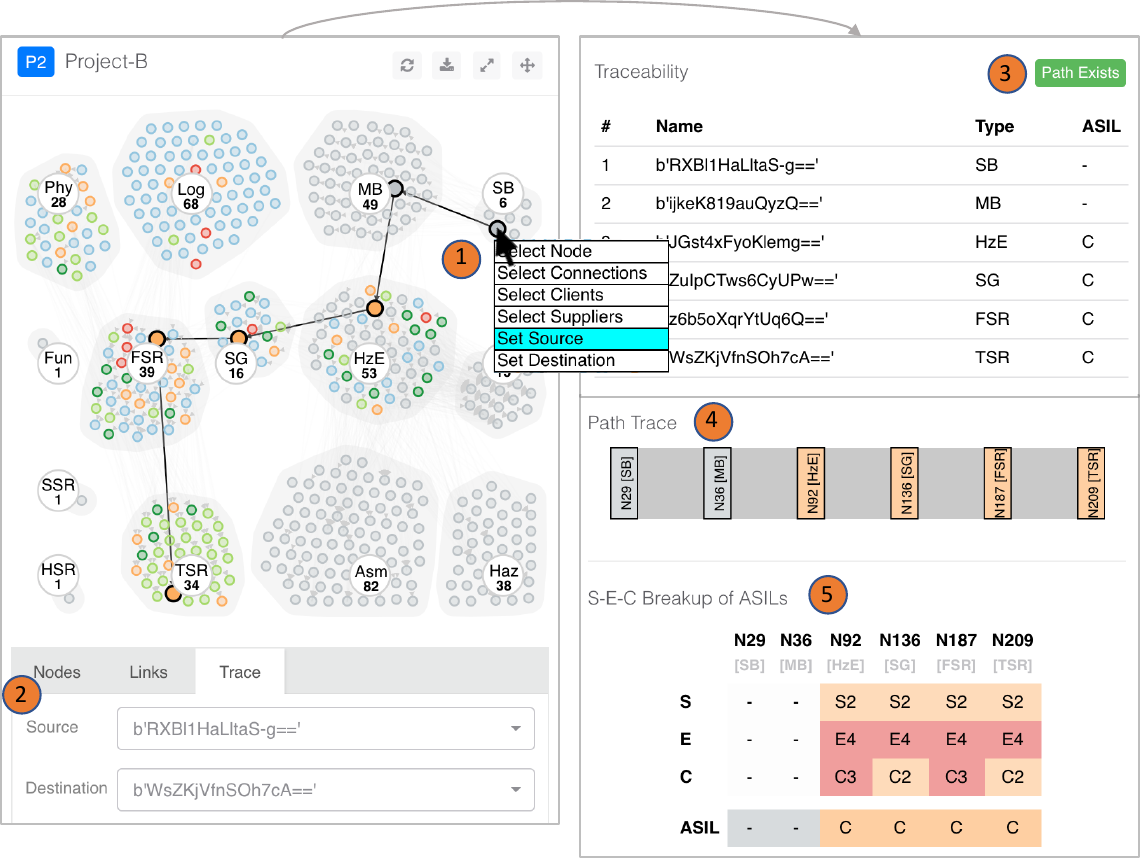}
        \caption{Finding if a path exists and trace the ASIL along it. \textcolor{orange}{\textbf{(1)}} Set a Source (\textit{System Behavior (SB)}) using the context menu (or search box) \textcolor{orange}{\textbf{(2)}} Set the Destination (\textit{Technical Safety Requirement (TSR)}) by looking it up in the search box (or context menu) \textcolor{orange}{\textbf{(3)}} Status of a path's existence along with its waypoints. \textcolor{orange}{\textbf{(4)}} A horizontal node-link diagram which helps visualize the Trace, as well as the ASILs of nodes. \textcolor{orange}{\textbf{(5)}} Heatmap to visualize the decomposition of ASIL ratings into \textit{Severity (S) - Exposure (E) - Controllability (C)}.}
        \label{fig:use-case-asil-trace}
    \vspace{-1.5em}
    \end{figure}

    \textbf{{Find if a Path exists between two Nodes}.}
    In the previous task, \userb traversed the network manually by following the nodes' connections. \app can also check if a path exists between non-adjacent nodes, that is, nodes that are not directly connected. Consider a scenario to check if a \emph{(Technical Safety Requirement (TSR)} exists for a \emph{System Behavior (SB)} as illustrated in Figure~\ref{fig:use-case-asil-trace}. \userb imports \projecta into the Compare View and switches to the Trace Tab in the Project View. \textcolor{orange}{\textbf{(1)}} They right click on the \emph{SB} and set it as Source from the context menu. \textcolor{orange}{\textbf{(2)}} They look up the \emph{TSR} from the search field and set it as Destination. \textcolor{orange}{\textbf{(3)}} \app calculates and finds that a path does exist between the two nodes. In this way, \app helps analysts find and visualize paths between two nodes.

    \textbf{Compare ASILs along a Path.}
    Figure~\ref{fig:use-case-asil-trace} also illustrates an extension of the previous task to analyze the ASILs of intermediary nodes along the path. \textcolor{orange}{\textbf{(4)}} The node-link (rectangles) diagram below reshapes the overlaid path in the visualization into a linear layout with the Source and Destination nodes at the ends. The rectangle's fill color is determined by the ASIL of the node. \userb observes that the same ASIL C has propagated from the \emph{Hazardous Event (HzE)} to the \emph{Technical Safety Requirement (TSR)}. To verify if the ASILs are really consistent, they need to analyze their respective breakups into Severity (S), Exposure (E), and Controllability (C) values. \textcolor{orange}{\textbf{(5)}} \app shows this decomposition of ASIL in the form of a heatmap. \userb observes that while \{N92, N136, N187, N209\} have consistent ASILs, they do not have consistent Severity (S), Exposure (E), and Controllability (C). For example, the Controllability of N136 is C2 whereas it should have ideally been C3 (from N92). In this way, \app helps analysts trace ASILs within the \fusashort network.

    \subsection{Find Common Elements Between Projects}
    Until now we have illustrated how \app helps \users explore and analyze one project at a time. \app also supports simultaneous comparative analysis among multiple projects. We illustrate this with another scenario.
    
    In Section \ref{exploration}, we illustrated how \app helped \usera find and report \emph{Malfunctioning Behaviors (MB)} that did not have a corresponding \emph{Hazardous Event (HzE)}. With resources likely already allocated for this project, \usera showed the list to his supervisor, \userc. On seeing the list, \userc is able to recall a previous project \projectc where the same elements may have already been defined. They open \app but this time select two projects: \projecta and \projectc from the Dashboard view. An overview (distributions and counts) of these projects including shared nodes and links can be seen from the Summary View similar to that shown in Figure~\ref{fig:summary-view}. \userc sees that there are common nodes and links among these projects and switches to the Compare View to find the specific ones.
    
\begin{figure}[t]
        \centering
        \includegraphics[width=1\columnwidth]{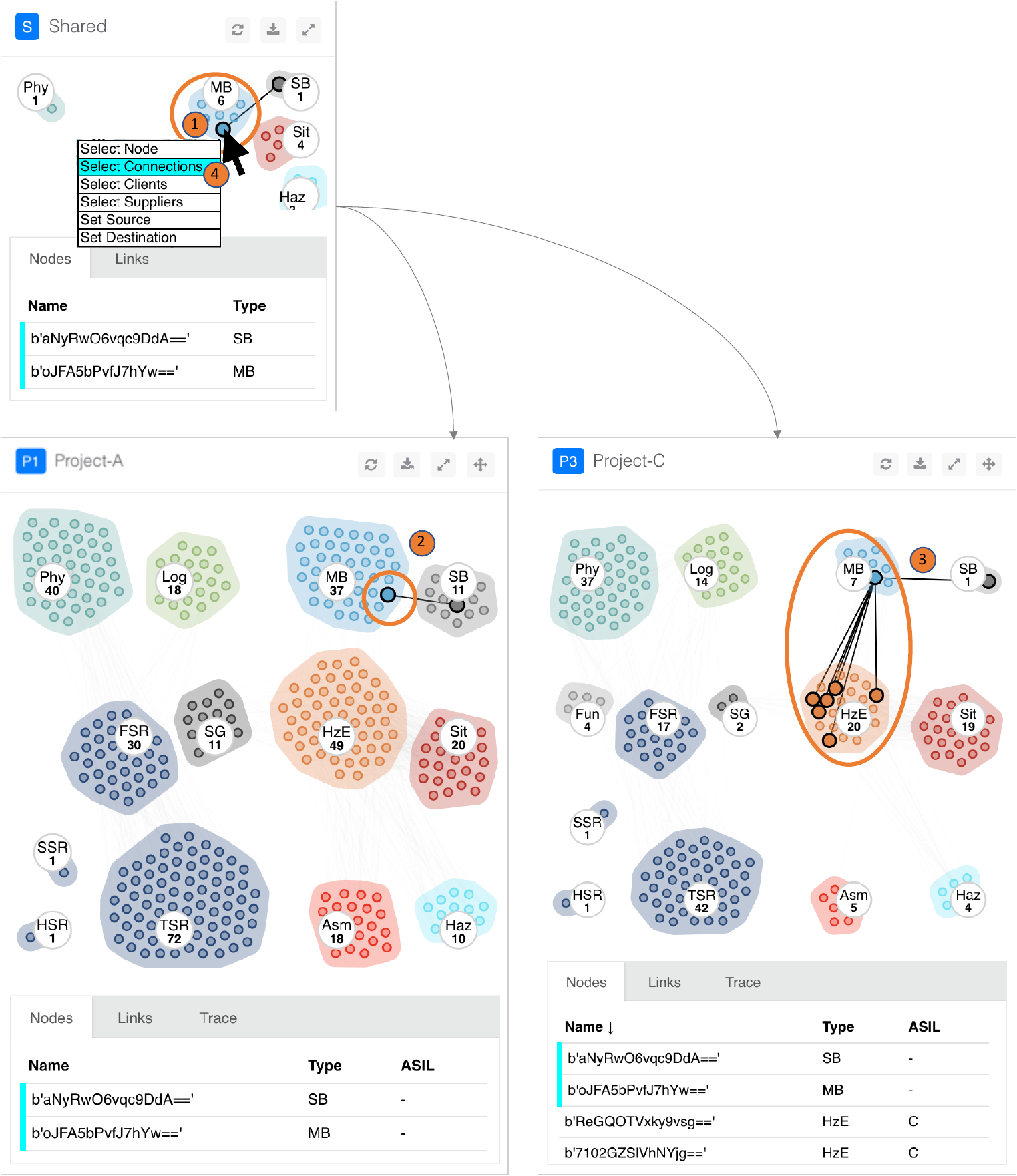}
        \caption{\textbf{Find common nodes among Projects}. \textcolor{orange}{\textbf{(1)}} Hover on a node in the Shared View highlights the common nodes and links in other project views. \textcolor{orange}{\textbf{(2)}} There is no link to \emph{HzE} in \projecta whereas \textcolor{orange}{\textbf{(3)}} \projectc has six links to \emph{Hze}. \textcolor{orange}{\textbf{(4)}} `Select(ing) Connections' from the context menu selects the nodes for further analysis.}
        \label{fig:use-case-comparative-analysis}
    \vspace{-1.5em}
    \end{figure}

    As illustrated in Figure~\ref{fig:use-case-comparative-analysis}, they see a new Shared Nodes panel at the top left. \app, by applying set intersection between network nodes and links, has pre-computed the shared nodes and links between the two projects and presented in the same node-link group visualization. They hover on a \emph{Malfunctioning Behavior (MB)} and find that it is in fact on \usera's list. \textcolor{orange}{\textbf{(1)}} Hovering on this node in the shared view highlights the corresponding nodes in \projecta and \projectc along with their connections, if any. While there are rightly no links from the \emph{MB} to any \emph{HzE} in \projecta \textcolor{orange}{\textbf{(2)}}, there are several links to \emph{HzEs} in \projectb \textcolor{orange}{\textbf{(3)}}. \userc right clicks the node of concern in \projectb and ``Selects Connections'' from the context menu \textcolor{orange}{\textbf{(4)}}. This selects all nodes that are connected to this node. \userc exports the list in \projectc and shares it with \usera who, instead of defining these connections from scratch, can just re-use them.

%% file: src/sections/discussion.tex
\subsection{Feedback}
\input{"src/sections/feedback.tex"}

\subsection{Deployment}
\input{"src/sections/deployment.tex"}

\subsection{Limitations}
\input{"src/sections/limitations.tex"}

%% file: src/sections/feedback.tex
To validate the design and functionality of \app, we gathered user feedback from the \revise{same domain experts with whom we conducted the initial feedback session and for whom the system was designed}. This was performed remotely, with our collaborator at the company demonstrating the prototype to users with their data (obfuscated in this paper). The overall feedback was positive.

Users commented that creating an overview encompassing several projects is currently very time consuming, and once more data is imported in to \app, would be something that can make their tasks more effective. Users also commented that seeing their data in \app opened up new questions and tasks.
For instance, one participant commented that \textit{``this let's me see projects quickly, and spot problems. But sometimes things don't look right, can I edit right here?''} \revise{Engineers currently edit the knowledge base that contains these files, but we will consider adding editing functionality as demonstrated in \cite{thomas_2006} and \cite{eichner_2016} in the future}.

Another participant asked \textit{``how do I only see what's changed in the visualization
from the last time I logged in? How do I know if this ASIL C has undergone revisions, e.g. was previously D?''} This opened up an entire aspect of visualizing temporal changes to data. This was fascinating as their database currently updates existing data with no provision to log the revision history. 
We will consider supporting this task of temporal evolution of ASIL in the future.

\revise{Few participants found aspects of the (node-link group) visualization challenging to follow initially and got more comfortable with use. For example, one participant asked, \textit{``what do these (so many node) colors represent?''}. Another participant asked, \textit{``what do the spatial positions of nodes mean?''}. This was expected considering our users' visualization literacy and will help us further improve the user experience.}

%% file: src/sections/deployment.tex
Given the positive feedback to date from pilot users, \app will be explored further. This process involves various data permissions, storage, and other infrastructure challenges. Once deployed, we plan to gather feedback from users.

%% file: src/sections/limitations.tex
\app in its current state has a few limitations. It is read-only, in that it does not support authoring data yet. It reads in data from the graph database only and does not support other ways such as manual file uploads. While it does have an export functionality to save the application state as an image, it does not support online collaboration features such as annotations, commenting, and sharing. We have tested \app to facilitate simultaneous comparison of up to five projects comprising close to thousand nodes and links each. The performance impact of visualizing more projects is yet to be tested.

%% file: src/sections/conclusion.tex
Automotive \users perform \fusashort of the vehicles to minimize risks associated with hazards. These \users currently spend a significant amount of time gathering and analyzing data. We designed and developed \app, a visual data analysis tool to help them visualize and interact with \fusa datasets. \app is a tool that aims to assist \fusa analysis by showing relationships and patterns across projects. This paper explains our design process, how the user tasks and design goals guided \app's interface design and development, and provided usage scenarios to explain how the tool can be used to perform existing as well newer advanced analysis tasks.